\documentclass[prb,twocolumn,showpacs]{revtex4}

\usepackage{graphicx}
\usepackage{dcolumn}
\usepackage{bm}

\begin{document}

\title{Frustrated Heisenberg Antiferromagnets between $d=2$ and $d=3$}

\author{A. Peles}\email{umpelesa@cc.umanitoba.ca}
\author{B.W. Southern}%
 \email{souther@cc.umanitoba.ca}
\affiliation{Department of Physics and Astronomy\\ University of Manitoba\\ Winnipeg, Manitoba\\ Canada R3T 2N2}%

\date{\today}

\begin{abstract}
The anitiferromagnetic Heisenberg model on a stacked triangular geometry with a finite number of layers is studied using
Monte Carlo methods. A topological phase transition occurs at finite temperature for all film thicknesses. Our results
indicate that topological excitations are important for a complete understanding of the critical properties of the
model between two and three dimensions.
\end{abstract}
\maketitle

The behavior of the Heisenberg antiferromagnet on triangular geometries 
is quite unusual. The ground state order parameter has SO(3) symmetry which
allows for a $Z_2$ topological defect.
In $d=2$,  there is a finite temperature vortex unbinding transition 
which is purely topological
in character\cite{kawmiya,kik,souxu,sou1,sou2,caffarel}.
Both above and below this
 temperature the spin-spin correlation function
decays exponentially with distance. This behavior is quite different 
from 
the  Kosterlitz-Thouless transition\cite{kosterlitz1,kosterlitz2}
 which occurs in two-dimensional $XY$ models. In the latter
case the correlations are exponential at high $T$ and decay with a power 
law at all temperatures below
the transition. There is no long range order but there is a finite spin 
wave stiffness in the low
temperature phase.   In the Heisenberg 
antiferromagnet, both the sublattice magnetizations
 {\it and} the spin wave stiffness are  zero
 at {\it all} finite temperatures. On the other hand
 the vorticity modulus is nonzero at low T and drops to zero at the transition. 
Hence the phase transition
is similar to the Kosterlitz-Thouless transition in that vortices 
are involved but it is different in that
the correlations decay exponentially at all finite temperatures. 

Field theoretic studies of this model in terms of
 the nonlinear
sigma model ($NL\sigma$) predict that there is only a transition at $T_c=0$ in two dimensions. 
However a fixed-dimension perturbative approach \cite{calabrese01,calabrese03} in $d=2$ using the corresponding Landau-Ginzburg-Wilson (LGW) Hamiltonian
has located a fixed point which may describe a topological phase transition.
In $d=3$ both theory and experiment 
indicate that there is a conventional phase
transition with the appearance of an order parameter below the critical temperature.
 However there
is a great deal of debate about the nature of this transition. Monte Carlo studies and perturbative field theory calculations 
in fixed dimension \cite{calabrese03b}
indicate a continuous transition
which belongs to a new chiral universality class whereas 
non-perturbative  RG (NPRG) studies indicate
 a phase
 transition which is very weakly first order with effective exponents \cite{tissl,NPRG}. 
A very recent numerical study of the RG flow in the $LGW$ Hamiltonian by Itakura \cite{Itakura} also suggests a  possible weak first order transition for the Heisenberg spin model but with much stronger evidence  for a first order transition in the case of  $XY$ spins 
\cite{peles03,peles03b}.  
The experimental studies of these three dimensional stacked triangular 
systems find
that they exhibit a continuous phase transition with a set of critical
 exponents that do not belong to any of known universality classes .

In this work we explore the crossover of the $d=2$ behavior to $d=3$. We  study the nature of 
the ordering
process of  a layered triangular system having dimensions $L \times L \times H$ with the
 number of stacked triangular layers ranging from $ H=2$ to $H=24$ and the linear size of each layer $L$ 
ranging from $18$ to $120$ by means of Monte Carlo
 simulations.

We consider an isotropic Heisenberg model of classical spins interacting via nearest 
neighbour
exchange on a layered triangular lattice described by
the following Hamiltonian
\begin{equation}
H = -\sum_{i<j} J_{ij} \vec{S}_{i} \cdot \vec{S}_{j} 
\label {eq:1}
\end{equation}
Within each triangular layer the
spins interact antiferomagnetically $J_{ij}<0 $. Between adjacent layers the 
interaction can be taken to be positive or negative
with no essential difference since there is no frustration associated with these interactions and hence there is 
a symmetry with respect to the sign.
 We shall choose it to be 
ferromagnetic $J_{ij}>0 $  and  set  $|J_{ij}|=J=1$ for all interactions.
Periodic boundary conditions are applied in the plane of  each triangular layer
 where, in order to preserve the $3$-sublattice
structure, it is important to choose $L$ to be a multiple of $3$. The finite layer 
thickness, $H$, can have any integer value since
we have chosen the inter-planar interactions to be ferromagnetic.
The top and bottom surface layers are subject to free boundary conditions.

The spin stiffness \cite{azaria3,p1}, or helicity modulus,
is a measure of the increase in free energy associated with a twisting of 
the order parameter in spin space by imposing
 a gradient of the twist angle about some axis $\hat {n}$
in spin space along some direction $\hat {u}$ in the lattice. 
Diagonal elements of the spin wave stiffness tensor can be calculated by choosing 
an orthogonal
triad for the  directions of   three principal axis in spin space. The  symmetry 
of the ground state  suggests that
two of the principal axes correspond to 
two perpendicular directions $\hat{\perp}_{1}$ and $\hat{\perp}_{2}$
 in the spin plane and that the third is perpendicular to this plane
 This third axis is  conveniently chosen  to point
along the average chirality direction $\hat{K}$. 

Another quantity of interest when topological properties of the system 
are to be considered is
the vorticity \cite{souxu}.  The vorticity is a measure of the response of the 
spin system to an imposed twist
about an given axis $\hat {n}$ in spin space along a closed path that encloses 
a vortex core.
The vorticity $V_{\hat{n}}$ contains a contribution due to the vortex core as well 
as part which is proportional to $\ln(L/a)$.
 This can be written as
\begin{equation}
V_{\hat{n}}= C_{\hat{n}}+v_{\hat{n}} \ln(L/a).
\label{ch4_vm}
\end{equation}
where $C_{\hat{n}}$ is a temperature dependent constant describing the core and   $v_{\hat{n}}$  is called
the the vorticity modulus. It plays a similar role to the spin stiffness and vanishes at a phase transition.

Azaria {\it et al.} \cite{azaria3} have made detailed predictions for the dependence of the stiffnesses 
on the linear size of the system
for single layers based on the continuum limit of this model. Their results for a 
nonlinear sigma model 
(NL$\sigma$)
using  renormalization group (RG) techniques  showed that the
spin wave stiffness of the triangular antiferromagnet is a nontrivial
 function of $\ln L$ at low temperatures
where $L$ is the linear dimension of the triangular lattice. At any finite 
temperature $T$, the stiffnesses are zero on length
scales large compared with the correlation length. However, on length 
scales $1 \ll L \ll \xi$, the stiffnesses are nonvanishing
at low $T$. 

\begin{figure} 
\centering
\includegraphics[width=3.5in,angle=0]{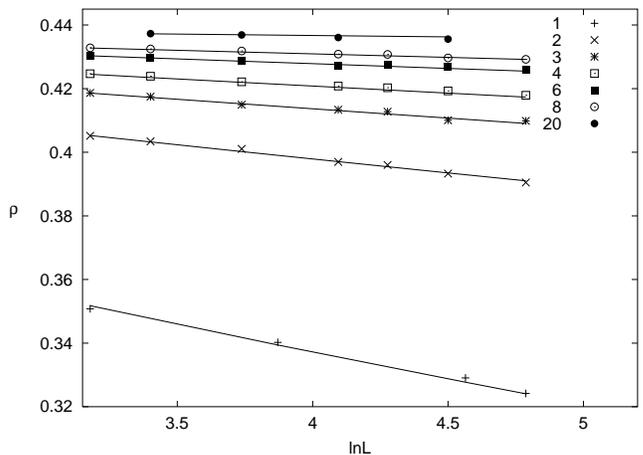}
\caption{Average spin stiffness  plotted as a function of $\ln L$ for different layer 
thicknesses $H$ at  $T=0.2$.The solid lines are the low temperature RG predictions.} 
\label{fig1}
 \end{figure}

In order to make a comparison of our 
calculated stiffnesses with the predictions of
the RG equations, we first calculate the stiffnesses for linear size $L=18$ so 
that the condition $H \ll L$ is satisfied.
Using the values at this length scale, we numerically integrate the expressions of Azaria {\it et. al.}
 with these initial  values 
to predict the behavior at larger length scales. We then compare the predicted 
curves with the values obtained directly using Monte
Carlo methods  at the relatively low temperature $T=0.2$ 
for different layer thicknesses. The dependence of the average of the three principal stiffnesses,
$\bar{\rho}=\frac{1}{3} \sum_{n=1}^{3}\rho_{n}$, on $\ln L$ is shown in figure 
\ref{fig1} for various values of $H$ .  In this figure the solid 
lines are the low temperature $T=0.2$ RG predictions
 and  the points are the Monte Carlo results. The data points agree very
 well with the predictions of the  NL$\sigma$ model at this
low temperature.  The stiffnesses decrease proportional to $\ln L$ and the results indicate that there is no stiffness at
large length scales. Hence it would appear that the finite layer systems behave in the 
same way as the single layer system with no
indication of a finite temperature phase transition in the stiffnesses.  However, as the 
the number of layers increases the slopes of
the lines in  figure \ref{fig1} decrease indicating that the stiffnesses 
should approach a finite value at large length
scales as the system becomes fully three dimensional.
 
\begin{figure} 
\centering
\includegraphics[width=3.5in,angle=0]{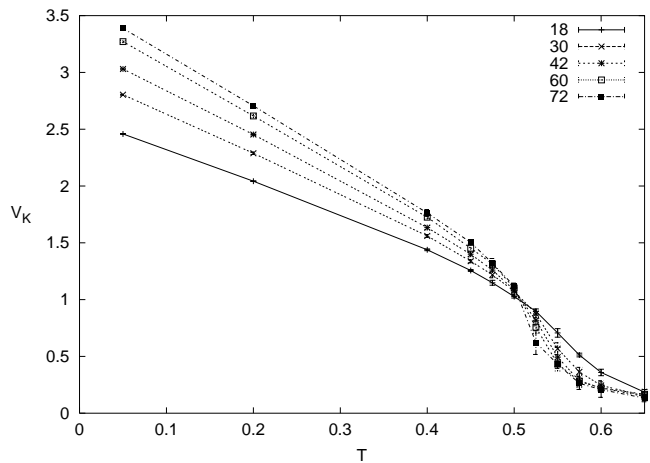}
\caption{Vorticity data for $H=2$ and different linear sizes $L= 18, 30, 42, 60, 72$ as function of temperature.} 
\label{fig2}
 \end{figure}

The excellent agreement of the Monte Carlo results and the RG predictions 
above indicate that topological
excitations are not important in this system at low $T$. However, at higher temperatures they may play an important
role. These topological degrees of freedom are not included in the RG analysis. In order to study
 the vortex degrees of freedom directly,
we have also calculated the response of the system to the presence of a single virtual vortex.
We calculate this vorticity response 
for a fixed number of layers $H$ and for various 
linear sizes $L$.  
In figure \ref{fig2} the temperature dependence of the vorticity 
data for $H=2$ is shown.
The curves for different sizes $L$ all cross at a common temperature point near $T=0.51$. This behavior 
suggests that there is a free energy cost
to creating isolated vortices in the system below this temperature but 
that they should spontaneously appear above it. 
At any value of $T$ we can study the size dependence of the vorticity 
using equation (\ref{ch4_vm}) to
extract the coefficient of the $\ln L$ term which we call a vorticity modulus. 
\begin{figure} 
\centering
\includegraphics[width=3.5in,angle=0]{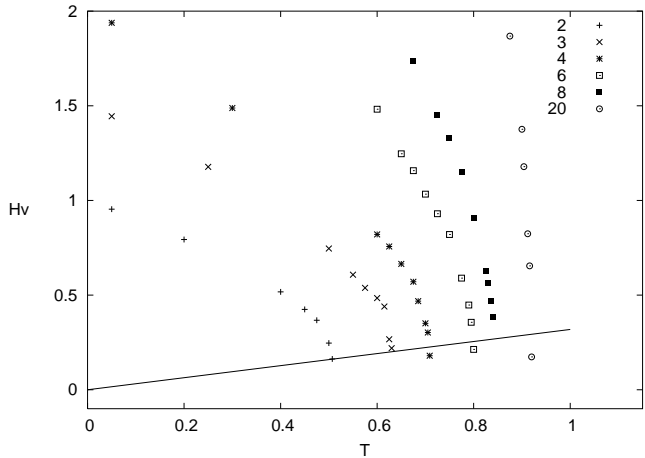}
\caption{Temperature dependence of $H$ times the average vorticity moduli 
for  $H=2, 3, 4, 6, 8$ and  $20$. The solid line represents the
ratio $Hv/T = \frac{1}{\pi}$.} 
\label{fig3}
 \end{figure}

The vorticity moduli
 have been determined for systems having a different number of layers $H$. 
In each case,   the  vorticity
 moduli abruptly drop to zero at the same temperature where the  vorticity curves cross.
Figure \ref{fig3} shows the average vorticity moduli 
$v = \frac{1}{3} \sum_{n=1}^3 v_n$ multiplied by the layer thickness $H$ 
as a function of temperature for several layer thicknesses. The solid line represents the ratio 
$Hv/T = \frac{1}{\pi }$ which intersects the
corresponding layer vorticity modulus at the temperature where the 
raw vorticity curves cross. This suggests that
the average vorticity modulus  exhibits a universal jump , $v/T_c=1/(\pi H)$, at the transition which
decreases as the number of layers $H$ increases.
\begin{figure} 
\centering
\includegraphics[width=3.5in,angle=0]{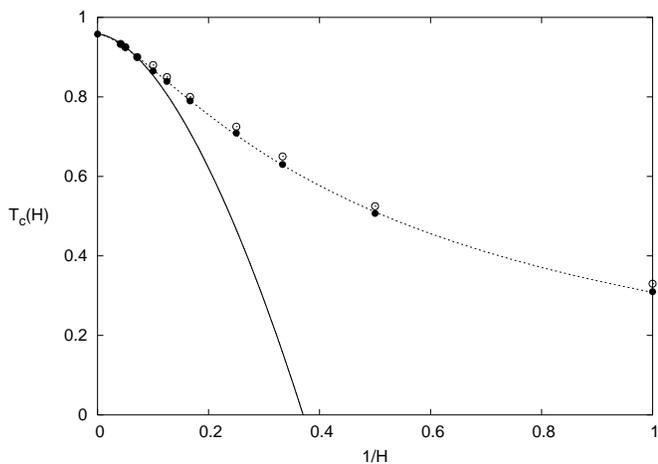}
\caption{$T_{c}(H)$ dependence on $1/H$ is indicated by the solid circles and the temperatures of the specific heat maxima
are indicated by open circles. The dotted line represents  a fit to
the prediction of FSS theory given in (\ref{fit}) and the solid line is the same fit with $H_1=0$. } 
\label{fig4}
 \end{figure}

Figure \ref{fig4} shows a plot of $T_{c}$ as well
as the temperature of the specific heat maximum versus the inverse number of layers . 
As  $\frac{1}{H} \rightarrow 0$  the
system becomes three dimensional with the critical temperature\cite{p1} equal to  $T_{c}(3d)=0.958$.
The specific heat maxima lie above the temperatures where the vorticity modulii drop to zero but the two
temperatures merge together as $1/H \rightarrow 0$.

The data points can be fit very well by a slightly modified form of the usual finite scaling (FSS)
\begin{equation}
T_{c}(H)=T_{c}(3d)\left(1-\frac{C_1}{(H+H_1)^{\frac{1}{\nu}}}\right)
\label{fit}
\end{equation}
where both $C_1$ and $H_1$ are constants. 
The correlation length exponent $\nu$ in the above expression is the value
 for the three dimensional case\cite{p1} and  equals $\nu$=  0.59.
Attempted fits of the data in figure \ref{fig4} (solid line) using this form  with $H_1=0$ are
are  only possible for large values of $H>8$ and there are
strong deviations for smaller values of $H$. 
The dotted line in figure \ref{fig4} corresponds to the values $C_1=7.28$ and $H_1=3.06$.
This modified form provides an excellent fit over the entire range
of layer thicknesses. 
\begin{figure}
\centering
\includegraphics[width=3.5in,angle=0]{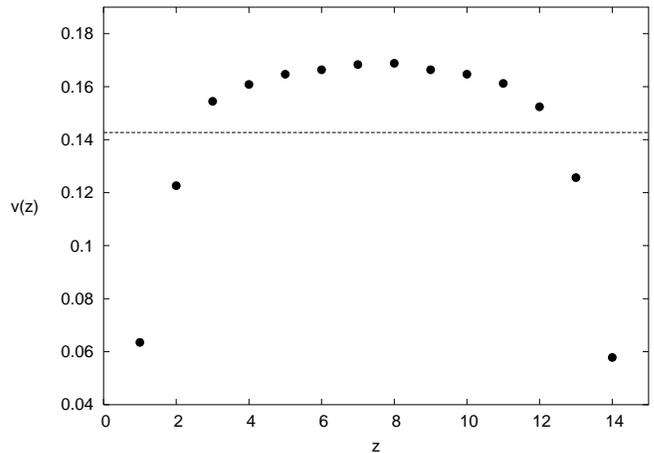}
\caption{Average vorticity $v(z)$ for $H=14$ for each individual layer at $T=0.8$. The dotted line is the average over all layers.} 
\label{fig5}
 \end{figure}

The quantity $H+H_1$ behaves as an effective layer width which may
be related to the boundary conditions at the upper and lower surfaces.
Figure \ref{fig5} show the average vorticity modulus $v(z)$ for each individual layer for the case of $H=14$ . The dotted horizontal curve  is the  average over all layers.  The deviations of $v(z)$
from the average are larger for thin films than for thicker films. For small values of $H$ the effect of surface layers is significant and in the
vicinity of the critical point the thickness $H$ is not the only relevant length scale.  It is being modified by the additional constant $H_1$ 
which is related to how  quickly $v(z)$ changes near the surface due to the free boundary conditions.

\begin{figure} 
\centering
\includegraphics[width=3.5in,angle=0]{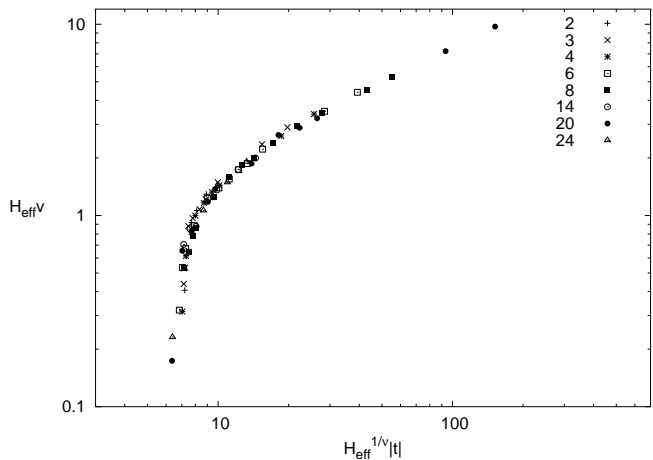}
\caption{FSS scaling plot of the average vorticity moduli $v$. } 
\label{fig6}
 \end{figure}

 Modifying finite scaling theory to include the effective width for small $H<8$ predicts that the vorticity moduli $v$ should scale with respect to the layer thickness $H$ as follows
\begin{equation}
(H+H_1)v=f((H+H_1)^{1/\nu}|t|).
\label{vormodsceff}
\end{equation}
where $\nu =0.59$ is the three dimensional correlation length exponent and $t$ is the reduced temperature $(T_c(3d)-T_c(H))/T_c(3d)$.
 Figure \ref{fig6} shows a plot of $H_{eff}v$ in terms of the variable $H_{eff}^{1/\nu} |t|$. The vorticity moduli curves for different thicknesses  collapse onto a universal curve. 

The results above are similar to those found in studies of  $^4$He superfluid fims\cite{ambeg,schultk1,schultk2}.
In these systems the order parameter is a complex number which has the same symmetry as the ferromagnetic $XY$
model.  The superfluid
transition in $d=2$ can be described in terms of the Kosterlitz-Thouless transition where vortices
destroy the ordered phase. Ambegaokar {\it et. al.}\cite{ambeg} argued that the KT theory can be extended to films of finite
thickness and that the transition should continue to have the $2d$ character provided the thickness is not too large.
Schultka and Manousakis\cite{schultk1} have studied the same problem numerically  and found that the results could indeed be
described by the KT theory using finite scaling theory. However, a detailed fit for a small number of layers required the
introduction of an effective width for the films.\cite{schultk2} 
Our results for the frustrated Heisenberg model are quite similar but there are some important differences. In the
$XY$ model, the transition is accompanied by a jump in the stiffness with a power law decay of the correlations below this temperature.
This behavior perists for films of arbitrary thickness until the $3d$ bulk limit is reached where the transition becomes the
usual $\lambda$ transition. In our case, the $d=2$ behavior observed\cite{kawmiya,souxu,sou1,sou2} for a single layer
also persists as the number of layers increases. The stiffness is zero at all finite $T$ indicating that the spin correlations
decay exponentially at all temperatures. However, the vorticity modulus exhibits a universal jump at a finite temperature
indicating that a topological phase transition occurs involving vortices. The size of the jump varies as $1/H$ and hence
approaches zero in the limit of $d=3$. In addition, the stiffness also approaches a nonzero value at temperatures below the
critical temperature in this same limit. This indicates that the $2d$ behavior approaches the $3d$ behavior in a
continuous way and that topological excitations play an important role between the two limits. The actual role that they play
in determining the nature of the transition in $3d$ is still not clear. It is possible that the Heisenberg model has a weak first
order transition as predicted by the NPRG  field theories and that the critical exponents are only
effective exponents. However, the evidence for first order behavior is much clearer for the $XY$ model and the
differences between Heisenberg and $XY$ behavior may in fact be due to the different types of topological defects. 

\begin{acknowledgments}
This work was supported by the Natural Sciences and Research Council of Canada and the HPC facility at the
University of Manitoba and HPCVL at Queen's University.
\end{acknowledgments}

\bibliography{layers}
\end{document}